
\input phyzzx
\input tables
\def\ts{\biggl(-{s \over 2}g_{\mu
\nu}+q_{2\mu}q_{1\nu}\biggr)}
\def\tut{\biggl({ut\over 2} \gm +up_\mu q_{1\nu}
              +sp_\mu p_\nu+tq_{2\mu}p_\nu\biggr)}

\def\ra{\rightarrow}

\def\tone{ q_{2\mu} p_\nu + p_\mu q_{1\nu}-q_{2\mu}
q_{1\nu}}
\def\ttwo{ t q_{2\mu} p_\nu - u p_\mu q_{1\nu}}

\def\gm{g_{\mu \nu}}
\def\tq{\biggl( {g^{\prime} \over g} \biggr)}
\def\frac#1#2{{#1 \over #2}}
\def\sp#1{{\rm Tr}\,\biggl(#1\biggr)}
\def\Sigmad{\Sigma^{\dagger}}

\def\refmark#1{[#1]}
\def\Tr{{\rm Tr}\,}
\PHYSREV

\frontpagetrue
{\baselineskip 14pt
\null
\line{\hfill FERMILAB-PUB-92/75-T}
\line{\hfill JHU-TIPAC-920009}
\vskip 1.0cm

\centerline{\bf EFFECTIVE FIELD THEORY OF}
\centerline{\bf ANOMALOUS GAUGE-BOSON COUPLINGS}
\centerline{\bf AT HIGH-ENERGY $pp$ COLLIDERS}

\vskip .75cm
\centerline{
{\bf J.~Bagger$^{(a)}$,}\foot{This work has been supported
by the U.S. National Science Foundation, grant
PHY-90-96198, and by the Alfred P. Sloan Foundation.}
{\bf  S.~Dawson$^{(b)}$}
\foot{This manuscript has been authored
under contract number DE-AC02-76-CH-00016 with the U.S.
Department of Energy.  Accordingly, the U.S. Government
retains a non-exclusive, royalty-free license to publish
or reproduce the published form of this contribution,
or allow others to do
so, for U.S. Government purposes.}
{\bf and G.~Valencia$^{(c)}$}}
\vskip .75cm
\centerline{$^{(a)}$ {\it Department of Physics and
Astronomy}}
\centerline{\it Johns Hopkins University, Baltimore,
MD~~21218}
\vskip 0.5cm
\centerline{$^{(b)}$ {\it Physics Department}}
\centerline{\it Brookhaven National Laboratory, Upton,
NY~~11973}
\vskip 0.5cm
\centerline{$^{(c)}$ {\it Theory Group}}
\centerline{\it Fermilab, Batavia, IL~~60510}
\vskip 1.75cm
\centerline{{\bf Abstract}}
\vskip 0.5cm

We compute the effects of anomalous gauge-boson
couplings at high-energy hadron colliders
using next-to-leading order $SU(2) \times SU(2)$
chiral perturbation theory.  By comparing the yields
from the universal $p^2$ terms with those arising from
new physics at order $p^4$, we estimate the sensitivity
of the SSC and LHC to the indirect effects of electroweak
symmetry breaking.

\vskip 0.25cm

\vfill
\line{Revised August 1992 \hfill}
\vfill

\chapter{Introduction}

\baselineskip20pt
\overfullrule0pt
The study of electroweak symmetry breaking is a major
reason for building new high energy hadron colliders.
At present, not much is known about the mechanism of
symmetry breaking, except for the fact that the
$W$ and $Z$ bosons have mass, or longitudinal
degrees of freedom.  This leads one to expect that
longitudinally-polarized gauge bosons will provide
an important probe of this new physics.

Over the past few years, the theory of electroweak
symmetry breaking has received much attention.  The case
that has been studied most is, of course, the standard
model with an elementary Higgs boson
\Ref\higgs{For a review of Higgs boson phenomenology,
see J.~Gunion, H.~Haber, G.~Kane and S.~Dawson, {\it The
Higgs Hunter's Guide}, (Addison-Wesley, Menlo Park, 1990).}.
If the Higgs boson is light, standard perturbative
calculations give a reliable indication of its prospects
for discovery.  If it is heavier than about 800 GeV,
however, the symmetry-breaking sector is strongly coupled
and naive perturbation theory breaks down
\REF\hhl{D.~Dicus and V.~Mathur, {\it Phys. Rev.} {\bf D7}
(1973) 3111.}
\REF\lqt{ B.~W.~Lee, C.~Quigg, and H.~B.~Thacker,
{\it Phys. Rev.} {\bf D16} (1977) 1519.}
\REF\hhlat{M.~L\"{u}scher and P.~Weisz, {\it Phys. Lett.}
{\bf 212B} (1989) 472.}\refmark{\hhl--\hhlat}.

The problem with a strongly-interacting symmetry-breaking
sector is that one cannot make firm predictions.  For
example, the strong dynamics might give rise to a resonance,
with the same quantum numbers as the standard-model Higgs
boson, whose properties are not simply related to the
parameters of the standard model
\REF\einhorn{M.~Einhorn, {\it Nucl. Phys.} {\bf B246}
(1984) 75; R.~Cahn and M.~Suzuki, {\it Phys. Rev. Lett.}
{\bf 67} (1991) 169.}\refmark{\lqt,\einhorn}.
Alternatively, the symmetry-breaking sector might not be
at all like the standard model, and many possibilities
have been suggested.  Typically, these theories give rise
to a rich spectrum of resonances in the TeV region, as
in technicolor models
\Ref\tc{E.~Farhi and L.~Susskind, {\it Phys. Rev.} {\bf D20}
(1979) 3404; {\it Phys. Rep.} {\bf 74} (1981) 277;
S.~Weinberg, {\it Phys. Rev.} {\bf D19} (1979) 1277;
E.~Eichten and K.~Lane, {\it Phys. Lett.} {\bf B90} (1980)
125.
A review of the current status of technicolor models
is given by B.~Holdom, {\it Model Building in Technicolor},
Lectures Given at the Nagoya Spring School, 1991, DPNU-91-
27.}.

On general grounds, one knows that some of the resonances
associated with electroweak symmetry breaking must couple
to the $V_L V_L$ final state.\foot{We generically denote the
longitudinal $W$ and $Z$ bosons by $V_L$.}
If the resonances are light enough to be produced at the
SSC or LHC, their discovery prospects depend on their
widths.  For example, the relatively narrow techni-rho
should be easy to see
\Ref\han{J.~Bagger, T.~Han
and R.~Rosenfeld, {\it Proceedings of the 1990
Snowmass Study on Physics in the 90's}, Snowmass, CO.}.
In contrast, a heavy Higgs is so broad that it
is difficult to isolate the signal from the
background \refmark{\higgs}.

Of course, it is also possible for the new
resonances to be too heavy to be produced at
the SSC or LHC.  In this case one would expect an
enhancement in the yield of $V_L V_L$ pairs.  Such
an enhancement would signal the onset of a new
strongly-interacting symmetry-breaking sector
\Ref\velt{M.~Veltman, {\it Acta Phys. Pol.} {\bf B8}
(1977) 475.},
with new resonances that lie out of reach
\REF\nlc{M.~Chanowitz, {\it Proceedings of the 23rd International
Conference on High Energy Physics}, Berkeley California,
(1986) edited by S.~Loken (World Scientific, Singapore, 1987).}
\REF\pedin{For a pedagogical introduction to the subject
see: M.~Chanowitz, {\it Electroweak Symmetry Breaking:
Higgs/Whatever}, Lectures presented at the SLAC Summer
Institute, Stanford California, 1989.}
\refmark{\nlc,\pedin}.

It is this second scenario that we will study in this paper.
We will use the formalism of effective field theory to
describe the symmetry breaking, and take the effective
Lagrangian to contain all the fields of the standard model
{\it except} the Higgs boson.  We will assume that the
resulting theory has an effective $SU(2) \times SU(2)$ chiral
symmetry, and replace the Higgs by an infinite set of
non-renormalizable operators whose coefficients parametrize
the low-energy effects of the electroweak symmetry breaking
\REF\appba{T.~Appelquist and C.~Bernard, {\it Phys. Rev.}
{\bf D22} (1980) 200.}
\REF\lon{A.~Longhitano, {\it Nucl. Phys.} {\bf B188} (1981)
118.} \refmark{\appba,\lon}.
In unitary gauge, these operators reduce to anomalous
couplings of the standard-model gauge bosons.

The formalism of effective Lagrangians provides a
well-defined computational framework for investigating
the physics of anomalous couplings and electroweak
symmetry breaking.  The
infinite set of terms in the effective Lagrangian
can be organized in an energy expansion.  At low
energies, only a finite number of terms contribute
to any given process.  At higher energies, more and
more terms become important, until the whole process
breaks down at the scale of the symmetry breaking
\Ref\cpt{S.~Weinberg, {\it Physica} {\bf 96A} (1979) 327;
J.~Gasser and H.~Leutwyler, {\it Ann. Phys.}
{\bf 158} (1984) 142;
{\it Nucl. Phys.} {\bf B250} (1985) 465.}.
A similar procedure gives an acceptable description of
$\pi \pi$ scattering amplitudes up to energies of about
$500$ MeV
\Ref\drv{J.~Donoghue, C.~Ramirez, and G.~Valencia,
{\it Phys. Rev.} {\bf D38} (1988) 2195.},
so we expect the effective Lagrangian
for $V_L V_L$ scattering to be reasonable up to
the TeV scale.

Thus, in this paper we shall study the sensitivity of
the SSC and LHC to new physics beyond the standard model.
We will assume that all new resonances lie out of reach,
and that the physics of electroweak symmetry breaking
is described by a model-independent low-energy effective
Lagrangian.  We will use this Lagrangian to carry out
a complete, order $p^4$, calculation of $V_LV_L$ pair
production in $pp$ colliders, for invariant masses
up to of order 1.0 TeV.  Our results extend
previous studies
\REF\wwcpt{A.~Dobado and M.~Herrero, {\it Phys. Lett.}
{\bf 228B} (1989) 495;
J.~Donoghue and C.~Ramirez, {\it Phys. Lett.}
{\bf B234} (1990) 361;
S.~Dawson and G.~Valencia, {\it Nucl. Phys.}
{\bf 352} (1991) 27.}
\REF\ramirthesis{C.~Ramirez, UMI-91-20931, Ph.D.~Thesis
(1991);
S.~Sint, Diplomarbeit Universit\"{a}t Hamburg (1991).}
\REF\simm{A.~Falk, M.~Luke, and E.~Simmons,
{\it Nucl. Phys.} {\bf B365 } (1991) 523.}
by including all initial states
and all next-to-leading corrections to the processes of
interest \refmark{\wwcpt--\simm}.

Throughout this paper we will use the electroweak
equivalence theorem to aid our analysis
\Ref\eqt{J.~M.~Cornwall, D.~N.~Levin, and G.~Tiktopoulos,
{\it Phys. Rev. } {\bf D10} (1974)1145; {\bf 11} (1975) 972
E; B.~W.~Lee, C.~Quigg, and H.~B.~Thacker, {\it Phys. Rev.}
{\bf D16} (1977) 1519; M.~S.~Chanowitz and M.~K.~Gaillard,
{\it Nucl.Phys.} {\bf B261} (1985) 379;
Y.-P.~Yao and C.~P.~Yuan, {\it Phys. Rev.}
{\bf D38} (1988) 2237;
J.~Bagger and C.~Schmidt, {\it Phys. Rev.}
{\bf D41} (1990) 264;
H.~Veltman, {\it Phys. Rev.} {\bf D41} (1990) 2294.}.
We will work in Landau gauge,\foot{In this gauge, the
would-be Goldstone bosons are massless and decouple
from the ghost sector.} and calculate our amplitudes
to leading order in $M^2_W/s$.  To any order in the
loop expansion, this amounts to keeping only those
terms of ``enhanced electroweak strength''
\Ref\smgbs{S.~Dawson and S.~Willenbrock,
{\it Phys. Rev.} {\bf D40} (1989) 2880;
M.~Veltman and F.~Yndurain,
{\it Nucl. Phys.} {\bf B325} (1989) 1.}.
Therefore our amplitudes are valid for the SSC and LHC,
but not for lower-energy machines.

We shall simplify our calculations by assuming factorization
of the production, scattering and decay of the $V_LV_L$
pairs.  For initial states with vector bosons, we will
use the effective $W$ approximation to compute
the luminosities of the transverse and longitudinal
polarizations.   We then fold these luminosities
with the scattering sub-processes to find the $pp$ cross
sections
\Ref\ewap{S.~Dawson, {\it Nucl. Phys.}
{\bf B249} (1985) 42; M.~Chanowitz and M.~Gaillard, {\it
Phys. Lett.}
{\bf B142} (1984) 85; G.~Kane, W.~Repko, and W.~Rolnick,
{\it Phys. Lett.} {\bf B148} (1984) 367.}.

In what follows we will assume that all of the new physics
associated with electroweak symmetry breaking is contained
in the vector boson self-couplings.  In particular, we
shall take the couplings of the fermions to be the same
as in the standard model, we will ignore the possibility
of extra pseudo-Goldstone bosons
\Ref\gluus{J.~Bagger, S.~Dawson, and G.~Valencia, {\it Phys.
Rev. Lett.} {\bf 67} (1991) 2256.}.
We will not discuss the
detection issues that go into analyzing the decay of
the longitudinal vector bosons, nor will we try to
define or study realistic experimental signatures.

\chapter{Effective Lagrangians}

\section{Global Symmetries}

In this paper we will assume that the
effective Lagrangian for electroweak symmetry breaking
is determined by new physics outside the reach of the
SSC or LHC.
Since we do not know the full theory, we must build the
effective Lagrangian out of all operators consistent
with the unbroken symmetries.  In particular, we must
include operators of all dimensions, whether or
not they are renormalizable.
In this way we construct the most general effective
Lagrangian that describes electroweak symmetry breaking.

To specify the effective Lagrangian, we must first fix the
pattern of symmetry breaking.  In the standard model, the
gauge group is $SU(2)_L \times U(1)_Y$, spontaneously
broken to the
$U(1)$ of electromagnetism.  The minimal global symmetry
consistent with this gauge group is $G = SU(2) \times U(1)$,
spontaneously broken to $H = U(1)$.  Of course, the global
symmetry group can also be larger.  For example, it could
be $G = SU(2) \times SU(2)$, broken to $H = SU(2)$, as
in the minimal standard model.  In this case, there is a
``custodial'' $SU(2)$ symmetry which ensures that $\rho=1$,
up to radiative corrections.\foot{In analogy with QCD, we
call the unbroken $SU(2)$ ``isospin.''}  Experimentally, we
know that $\rho \simeq 1$, so we will adopt the second group,
and assume that the custodial $SU(2)$ symmetry\foot{Note
that this is a very strong
assumption.  The constraint $\rho \simeq 1$ affects
only one term in the full effective Lagrangian.}
is broken {\it only} by terms that vanish as the hypercharge
coupling\foot{The custodial $SU(2)$ symmetry is also
broken by the mass splittings in the fermion doublets.
We shall ignore this symmetry breaking in what follows.}
$g'\rightarrow 0$.

Let us start by constructing the effective Lagrangian
associated with breaking $SU(2) \times SU(2) \rightarrow
SU(2)$.  We introduce the would-be Goldstone boson fields
$w^+$, $w^-$ and $z$, as well as the gauge fields $W_i$
and $B$, through the matrices
$$
\eqalign{
\Sigma\ \equiv&\ \exp{\biggl( {i w^i \tau^i \over v}
\biggr)} \crr
W_\mu\ \equiv&\  W^i_\mu \tau^i \crr
B_{\mu \nu}\ =&\ {1 \over 2}\biggl(\partial_\mu B_\nu
-\partial_\nu B_\mu \biggr) \tau^3 \crr
W_{\mu \nu}\ =&\ {1 \over 2} \biggl(\partial_\mu W_\nu -
\partial_\nu
W_\mu
-{i\over 2}\,g [W_\mu,W_\nu]\biggr)\ , \cr}
\eqn\cov
$$
where the $\tau^i$ are Pauli matrices, normalized so that
${\rm Tr}(\tau^i \tau^j ) =2 \delta^{ij}$.  The derivative
$$
D_\mu \Sigma \ =\ \partial_\mu \Sigma\ +\ {i \over 2}\,g
W_\mu
\Sigma \ - \ {i \over 2}\,g^{\prime} B_\mu \Sigma \tau^3
\eqn\covd
$$
transforms covariantly under global $SU(2) \times SU(2)$
transformations,
$$
\Sigma\ \rightarrow\  L\, \Sigma\, R^{\dagger}\ ,
\eqn\trans
$$
and $g$ and $g^{\prime}$ are the  coupling constants
of the gauged $SU(2)_L$ and $U(1)_Y$ respectively.

The lowest-order term in the effective Lagrangian
contains two derivatives,
$$
{\cal L}^{(2)}\ =\ {v^2 \over 4}\, {\rm Tr}\,
D^{\mu}\Sigma^{\dagger}
D_{\mu}\Sigma\ .
\eqn\lola
$$
The couplings of the would-be Goldstone bosons to the
$SU(2)_L \times U(1)_Y$ gauge fields are fixed by the
covariant derivative.  To this order, the effective
Lagrangian is unique.  The full Lagrangian is the sum
of the lowest-order effective Lagrangian, together
with the usual gauge-boson kinetic energy,
gauge-fixing and Fadeev-Popov terms.

The next-to-leading order  terms in the effective Lagrangian
contain six free parameters:\foot{We have normalized the
coefficients as in Ref.~\simm, so they are
``naturally'' ${\cal O}(1)$.  Note that we have
included one term that is formally of order $p^2$.
This term is induced at one loop by hypercharge
gauge-boson exchange.  Therefore $\Delta\rho$
is proportional to $g^{\prime2}/16\pi^2$, and
the term can be considered to be of order $p^4$.}
$$
\eqalign{
{\cal L}^{(4)}\ =&\  {L_1 \over 16 \pi^2}\, \biggl[
\sp{D^\mu\Sigmad D_\mu \Sigma} \biggr]^2
\ +\  {L_2 \over 16 \pi^2}\, \sp{D_\mu\Sigmad D_\nu \Sigma}
\sp{D^\mu\Sigmad D^\nu \Sigma} \crr
\ &-\ i g {L_{9L} \over 16 \pi^2}\, \sp{W^{\mu \nu} D_\mu
\Sigma D_\nu \Sigmad}
\ -\ i g^{\prime} {L_{9R} \over 16 \pi^2}\, \sp{B^{\mu \nu}
D_\mu \Sigmad D_\nu\Sigma} \crr
\ &+\ g g^{\prime} {L_{10}\over 16 \pi^2}\, \sp{\Sigma
B^{\mu \nu}
\Sigmad W_{\mu \nu}}
\ +\  {1 \over 8}\, \Delta\rho\, v^2 \,\biggl[\sp{\tau^3
\Sigmad
D_\mu\Sigma}\biggr]^2\ .
\cr}
\eqn\nela
$$
To this order, these are the only terms when $G = SU(2)
\times SU(2)$, broken only by the hypercharge coupling $g'$.
One can
think of other terms, such as $\Tr(D^2 \Sigmad D^2 \Sigma),$
but they can all be absorbed in Eq.~\nela\ by using the
lowest-order equations of motion, $\Sigma D^2 \Sigmad =
(D^2 \Sigma) \Sigmad$.  Therefore we construct our
${\cal O}(p^4)$ amplitudes
by using Eq.~\lola\ at the tree and one-loop levels, and
Eq.~\nela\ at tree level {\it only}.

\section{Renormalization Scheme}

As usual with effective Lagrangians,
we renormalize our amplitudes using a mass-independent
renormalization scheme.  To order $p^4$, the infinities
that appear at one loop can all be absorbed by defining
renormalized parameters $L^r_i(\mu)$.  Therefore we use
dimensional regularization and adopt the following
renormalization scheme:
$$
\eqalign{
L_1^r(\mu)\ =&\ L_1+{1 \over 24}
\biggl({1 \over \hat{\epsilon}} + {5 \over 3}\biggr)\crr
L_2^r(\mu)\ =&\ L_2+{1 \over 12}
\biggl({1 \over \hat{\epsilon}}+ {13 \over 6}\biggr)\crr
L_{9L}^r(\mu)\ =&\ L_{9L}+{1 \over 12}
\biggl({1 \over \hat{\epsilon}}+{8 \over 3}\biggr)\crr
L_{9R}^r(\mu)\ =&\ L_{9R}+{1 \over 12}
\biggl({1 \over \hat{\epsilon}}+{8 \over 3}\biggr)\crr
L_{10}^r(\mu)\ =&\ L_{10}-{1 \over 12}
\biggl({1 \over \hat{\epsilon}}+{8 \over 3}\biggr)\ ,
\cr}
\eqn\renl
$$
where
$$
{1 \over \hat{\epsilon}}\ =\ {2 \over 4-n} - \gamma
+\log(4\pi)
-\log(\mu^2)\ .
\eqn\eppe
$$
These definitions remove extraneous constants that can be
absorbed into redefinitions of the $L^r_i(\mu)$ in our
amplitudes.

As mentioned earlier, we work to lowest order in the
electroweak couplings, and compute the leading corrections
of enhanced electroweak strength.  Effectively this
means that when we compute one-loop diagrams, we allow
{\it only} would-be Goldstone bosons in the loops.  This
implies that we do not need to renormalize the usual
gauge-boson sector of the theory.

To one-loop order, the $\Delta\rho$ term in the effective
Lagrangian is renormalized by diagrams with a hypercharge
gauge boson in the loop.  It is not renormalized at all
if we only consider the terms of enhanced electroweak
strength.\foot{Recall that we assume $H
= SU(2)$.}  Therefore, we do not need to
specify a renormalization for the
coefficient $\Delta\rho$.

For studies at energies on the order of the
$W$ mass, it is not possible to separate the terms into
those of electroweak and enhanced electroweak strength.
One must calculate beyond leading order in $g$ or $g'$,
and introduce a renormalization scheme for the usual gauge
sector of the electroweak interactions.  One also needs an
additional counterterm for $\Delta\rho$.

All these complications become necessary in studies for
lower energy machines, such as LEP2 and the Tevatron.
We are able to avoid these issues because we concentrate
on a kinematic regime where the only relevant terms are
those of enhanced electroweak strength.

\section{Present Constraints}

We can gain some insight into the constraints on
our $SU(2) \times SU(2)$ effective Lagrangian by reducing
it to unitary gauge, with $\Sigma =1$.  In this gauge
the new physics appears in the form of anomalous
gauge-boson couplings.

Let us first consider the term with $L_{10}$, which reduces to
$$
\qquad
{g g^\prime \over 2} {L_{10} \over 16 \pi^2}
\biggl( \partial_\mu B_\nu -\partial_\nu B_\mu \biggr)
\biggl[ \partial_\mu W^3_\nu - \partial_\nu W^3_\mu
\ -\ ig(W^+_\mu W^-_\nu\ -\ W^-_\mu W^+_\nu)\ \biggr]
\ .
\eqn\lpss
$$
This term contains a three-gauge-boson coupling, as well as an
``oblique'' correction to the gauge-boson self-energies
\Ref\obcl{B.~Holdom and J.~Terning,
{\it Phys. Lett.} {\bf 247B} (1990) 88; M.~Golden and
L.~Randall,
{\it Nucl. Phys.} {\bf B361} (1991)  3; A.~Dobado, D.~Espriu
and M.~Herrero, {\it Phys. Lett. } {\bf 225B} (1990) 405.}.
This correction is related to the $S$ parameter that occurs
in electroweak radiative corrections for $Z$-physics,
$L^r_{10}(M_Z) = -\pi S$
\REF\pes{D.~C.~Kennedy and B.~W.~Lynn, {\it Nucl. Phys.}
{\bf B322} (1989) 1;
M.~Peskin and T.~Takeuchi, {\it Phys. Rev. Lett.} {\bf 65}
(1990) 964.}
\REF\mar{W.~Marciano and J.~Rosner, {\it Phys. Rev. Lett.}
{\bf 65} (1991) 2963.}
\refmark{\pes,\mar}.
A recent best fit to the data gives $L^r_{10}(M_Z) = 0.0 \pm 1.6$
\Ref\altarelli{G.~Altarelli, R.~Barbieri and S.~Jadach, {\it
Nucl. Phys.} {\bf B369} (1992) 3; G.~Altarelli CERN-TH 6525/92.},
which we translate into
$$
L^r_{10}(\mu)\ =\ 0.5\ \pm\ 1.6\ ,
$$
for $\mu = 1.5$ TeV.

\REF\scott{ K.~Hagiwara, R.~Peccei, D.~Zeppenfeld and
K.~Hikasa, {\it Nucl. Phys.} {\bf B282} (1987) 253;
D.~Zeppenfeld and S.~Willenbrock, {\it Phys. Rev. }
{\bf D37} (1988) 1775.}

\REF\sntgb{E.~Yehudai, SLAC-Report 383, Ph.D.~Thesis (1991),
and references therein.}

\REF\hgb{
B.~Holdom, {\it Phys. Lett.} {\bf 258B} (1991) 156.}

The terms with $L_9$ contain anomalous three-
and four-gauge-boson couplings.  The standard notation for
anomalous three-gauge-boson couplings is given by
\refmark{\scott,\sntgb,\hgb}
$$
 -\ ie\, \kappa_\gamma\, W^+_\mu W^-_\nu A^{\mu \nu}
\ -\ ie\, {\rm cot}\theta\, \kappa_Z \, W^+_\mu W^-_\nu
Z^{\mu \nu}\ ,
\eqn\anomtgg
$$
where \refmark{\simm,\hgb}
$$
\kappa_Z - 1 \ \simeq\ \kappa_\gamma - 1 \ \equiv\ \Delta\kappa
\ \simeq\ {\cal O} \biggl( g^2\, { L_{9 L,R} \over 16 \pi^2
}
\biggr)\ .
\eqn\relkl
$$
\REF\kvy{G.~Kane, J.~Vidal and
C.-P.~Yuan, {\it Phys. Rev.} {\bf D39} (1989) 2617.}
An analysis by Kane, Vidal and Yuan \refmark{\kvy}
finds that the SSC will be sensitive to values of
$|\Delta \kappa| \gsim 0.15$, which we translate to
$|L^r_{9L,R}(\mu)| \gsim 25$, for $\mu \simeq$ 1.5
TeV.\foot{We interpret the results of Refs. \simm\
and \kvy\ in terms of running couplings evaluated
at $\mu \simeq 1.5$ TeV.}
This is compatible with the results of Falk, Luke and
Simmons \refmark{\simm}, who studied $pp\rightarrow W^{\pm}
Z$ and $pp\rightarrow W^{\pm} \gamma$ and found that the
SSC will {\it not} be sensitive to the values
$$
\eqalign{
-16\ \lsim &\ L^r_{9L}(\mu)\ \lsim\ 7 \cr
-119\ \lsim &\ L^r_{9R}(\mu)\ \lsim\ 113\ ,}
\eqn\xxxxx
$$
evaluated at $\mu \simeq$ 1.5 TeV.
The present bounds are of order
\Ref\samuel{UA2 Collaboration, J.~Alitti {\it et. al.}, {\it Phys. Lett.}
{\bf 277B} (1992) 194;
M.~Samuel, G.~Li, N.~Sinha, R.~Sinha,
M.K.~Sundaresan, {\it Phys. Rev. Lett.}
{\bf 67} (1991) 9; (E) 2920.}
$$
-2.2\ \leq\ \kappa_\gamma-1\ \leq\ 2.6 \ ,
\eqn\yyyyyy
$$
although the Tevatron is expected to reach a
sensitivity of about 1.3 for $|\kappa_\gamma-1|$
\REF\lay{J.~Layssac, G.~Moultaka, F.~Renard, G.~Gounaris,
Preprint PM-90-42 (1991)}
\REF\arg{E.~Argyres, O.~Korakianitis, C.~Papadopoulos,
W.~Stirling, {\it Phys. Lett.} {\bf B259} (1991) 195.}
\Ref\cdfb{U.~Baur and E.~Berger, {\it Phys. Rev.} {\bf D41}
(1990) 1476.}.
This is similar to the expected sensitivity of
LEP2 \refmark{\scott,\kvy,\lay,\arg}.

\FIG\fiun{Values of $L^r_1(\mu)$ and $L^r_2(\mu)$ at
$\mu=1.5$ TeV allowed by the unitarity constraints
described in the text. The solid
(dashed) line is the boundary of the allowed
region when one requires the partial waves
with $J \leq 2$ to satisfy $|Re(a^I_J)|<1/2$
for $M_{VV}\leq 1.0$ TeV  ($M_{VV}\leq 1.5$ TeV).}

The terms with $L_1$ and $L_2$ give rise to
anomalous four-gauge-boson couplings.
These couplings are not
constrained by experiment.  They are limited,
however, by perturbative unitarity.
As we will see in the following sections, $L^r_1(\mu)$
and $L^r_2(\mu)$ contribute to the $V_LV_L \ra V_LV_L$
scattering amplitudes.  These amplitudes grow with energy,
which allows us to place ``unitarity bounds'' on the
parameters.  The procedure is simple:  one first computes
the partial wave amplitudes $a^0_0$, $a^2_0$, $a^1_1$,
$a^2_0$ and $a^2_2$ to one loop, and then demands that
they do not violate the elastic partial wave unitarity
condition $|Re(a^I_J)|\leq 1/2$ below
$M_{VV}=1.0$ TeV (or $M_{VV}=1.5$ TeV). This gives the
``allowed'' regions shown in Fig.~\fiun. By considering
inelastic unitarity
constraints on the $q\overline{q}\ra V_L V_L$ amplitudes,
one can
perform a similar exercise to bound $L^r_{9L}(\mu)$ and
$L^r_{9R}(\mu)$ .
It turns out, however, that this does not
significantly constrain $L^r_{9L,R}(\mu)$.

The final term in \nela\ is proportional to
$\Delta \rho$.  By power counting, we
assign a factor of
$g^{\prime 2}/16\pi^2$ to this coupling,\foot{As
mentioned before, this term arises at one loop
from hypercharge gauge-boson exchange.  It is
not renormalized by loops containing only
would-be Goldstone bosons.  Since we are
only computing the corrections of enhanced
electroweak strength,
any ``bare'' value of $L_{12}$ is
not renormalized, and $L^r_{12}$ is
independent of $\mu$.}
$$
\Delta\rho\ \equiv\ g^{\prime 2}\,{  L^r_{12}
 \over 16 \pi^2 } \ .
\eqn\zzzzzzz
$$
Combining LEP data and low energy data from deep
inelastic scattering and parity violation in Cesium,
Altarelli \refmark{\altarelli}
finds $\Delta\rho = 0.0016\pm 0.0032$, which gives
$L^r_{12} \simeq 2.0 \pm 4.0$.  We shall
set $\Delta\rho = 0$ in what follows.

With these counting rules, all other terms
in Ref.~\lon\ with four derivatives
are actually of order $p^6$.
In particular, this includes a term
recently discussed by Holdom
\Ref\hez{B.~Holdom, {\it Phys. Lett.}
{\bf 259B} (1991) 329.},
$$
K\, \sp{\tau_3\Sigmad D_\mu D_\nu \Sigma}
\sp{\tau_3(D_\mu D_\nu \Sigmad) \Sigma}\ .
\eqn\othersbt
$$
This term is related to the observable $U$
that has been used in the study of electroweak
radiative corrections.  Other anomalous
three-gauge-boson couplings considered in the
literature include\foot{Note that power counting
indicates that these terms should not be suppressed
by $M^2_W$, but by $\Lambda^2 \lsim 16 \pi^2 v^2$.}
$$
{\lambda_\gamma \over M^2_W}\ W^+_{\lambda \mu}
W^{-\mu}_\nu A^{\nu
\lambda}\ +\
{\lambda_Z \over M^2_W}\ W^+_{\lambda \mu} W^{-\mu}_\nu
Z^{\nu
\lambda}\ .
\eqn\othertgb
$$
These couplings are of order $p^6$ and are
suppressed within the framework of our discussion.

\section{Typical Coefficients}

\REF\valdr{J.~Donoghue, C.~Ramirez and G.~Valencia, {\it
Phys. Rev.} {\bf D39} (1989) 1947.}

In order to understand the bounds on the coefficients
$L^r_i(\mu)$, it is instructive to estimate the size
of the coefficients in typical theories.  Using the
effective Lagrangian approach, this can be done in a
consistent way.  We first consider a model with
three would-be Goldstone bosons, interacting with
a scalar, isoscalar resonance like the Higgs boson.  We
assume that the $L^r_i(\mu)$ are dominated by tree-level
exchange of the scalar boson.  If one integrates out the
scalar, and matches coefficients at the scale $M_H$, one finds
\refmark{\lon,\cpt,\valdr}:
$$\eqalign{
L^r_1(\mu)\ =&\ {64 \pi^3 \over 3}\,
{\Gamma_H v^4 \over M^5_H}\ +\
{1 \over 24}\log\biggl({M^2_H \over \mu^2} \biggr)
\crr
L^r_2(\mu)\ =&\ {1 \over 12}\log\biggl({M^2_H \over \mu^2}
\biggr)
\crr
L^r_{9L}(\mu)\ =&\ L^r_{9R}(\mu)\ =\
{1 \over 12}\log\biggl({M^2_H \over \mu^2} \biggr)
\crr
L^r_{10}(\mu)\ =&\ -{1 \over 12}\log\biggl(
{M^2_H \over \mu^2} \biggr)\ ,
\cr}
\eqn\lsmn
$$
where $\Gamma_H$ is the width of the scalar into
Goldstone bosons.  If we naively take
$$
\Gamma_H\ =\ {3 M^3_H \over 32 \pi v^2}\ ,
\eqn\gammah
$$
as in the standard model, we find the values quoted in
Table 1, assuming $\mu = 1.5$ TeV.

\topinsert
\centerline{{\bf Table 1}}
\centerline{Coefficients induced by a scalar,
isoscalar particle like the Higgs,}
\centerline{with $\mu = 1.5$ TeV.}
\vskip.3in
\begintable
\hfil \quad $M_H$ (TeV) \quad \hfil |
\hfil \qquad $L^r_1(\mu)$ \qquad \hfil |
\hfil \qquad $L^r_2(\mu)$ \qquad \hfil |
\hfil \qquad $L^r_9(\mu)$ \qquad \hfil |
\hfil \qquad $L^r_{10}(\mu)$ \qquad \hfil \crthick
2.0 | 0.33 | 0.01 | 0.01 | $-0.01$ \crthick
1.5 | 0.55 | 0.00 | 0.00 |  0.00
\endtable
\bigskip
\vfill
\endinsert

\REF\ramirezz{C.~Ramirez, {\it Phys. Rev.} {\bf D42} (1990)
1726.}
\REF\veltmann{H.~Veltman and M.~Veltman, {\it Acta Phys.
Pol.} {\bf B22}
(1991) 669.}

Let us now consider a second model for the $L^r_i(\mu)$,
and assume that the coefficients are dominated by
tree-level exchange of a
rho-like particle with spin and isospin one.
Integrating out the rho, and matching coefficients at the
scale $M_\rho$, one finds:\foot{See the second paper of
Ref.~\cpt.  The contribution to $\Delta\rho$
is computed in Ref.~\ramirezz.  Related issues
are discussed in Ref.~\veltmann.}
$$
\eqalign{
L^r_1(\mu)\ =&\ {1 \over 24}\biggl[-{96 \pi^2 f^2 \over
M^2_{\rho}}
\ +\ \log\biggl({M^2_\rho \over \mu^2} \biggr)
 \biggr] \crr
L^r_2(\mu)\ =&\ {1 \over 12}\biggl[{48 \pi^2 f^2 \over
M^2_{\rho}}
\ +\ \log\biggl({M^2_\rho \over \mu^2} \biggr)
 \biggr] \crr
L^r_{9L}(\mu)\ =&\ L^r_{9R}(\mu)\ =\
{1 \over 12}\biggl[{96 \pi^2 f F_{\rho} \over M^2_{\rho}}
\ +\ \log\biggl({M^2_\rho \over \mu^2} \biggr)
 \biggr] \crr
L^r_{10}(\mu)\ =&\ -{1 \over 12}\biggl[{48 \pi^2 F_{\rho}^2
\over
M^2_{\rho}}
\ +\ \log\biggl({M^2_\rho \over \mu^2} \biggr)
 \biggr] \cr}
\eqn\terhn
$$
where the constant $f$ is related to the width
$\Gamma_\rho$,
$$
\Gamma_{\rho}\ =\ {1 \over 48 \pi}{f^2 \over
v^4}M^3_{\rho} \ ,
\eqn\intwm
$$
and $F_{\rho}$ is defined by:
$$
\langle 0|V^i_\mu |\rho^k(p)\rangle \ =
\ \delta^{ik} \epsilon_\mu F_{\rho} M_{\rho}\ .
\eqn\intwmm
$$
To estimate these parameters, we will model the resonance
by a techni-rho whose properties are fixed by the
ordinary QCD rho.  Using large-$N$ scaling arguments for
the mass and width of the resonance, we find the values
quoted in Table 2, for $\mu = 1.5$ TeV.

\topinsert
\centerline{{\bf Table 2}}
\centerline{Coefficients induced by a vector, isovector
particle like the techni-rho,}
\centerline{with $\mu = 1.5$ TeV.}
\vskip.3in
\begintable
\hfil \quad $M_\rho$ (TeV) \quad \hfil |
\hfil \qquad $L^r_1(\mu)$ \qquad \hfil |
\hfil \qquad $L^r_2(\mu)$ \qquad \hfil |
\hfil \qquad $L^r_9(\mu)$ \qquad \hfil |
\hfil \qquad $L^r_{10}(\mu)$ \qquad \hfil \crthick
2.0 | $-0.31$ | 0.38 | 1.4 | $-1.5$ \crthick
1.5 | $-0.60$ | 0.60 | 2.4 | $-2.5$
\endtable
\bigskip
\vfill
\endinsert

In either case, the $L^r_i(\mu)$ are numbers of order one,
which implies that $\Delta\kappa \simeq {\cal O}
(g^2/16\pi^2)$.
Values much larger than these are associated with
light resonances (or other light particles not present
in the minimal standard model), in which case chiral
perturbation theory breaks down at a very low energy.
Since one would presumably see the new particles directly,
it does not make sense to take the $L^r_i(\mu)$
much larger than one.

\FIG\fgbs{Counting rules for the production
of $V_L V_L$ pairs through vector boson fusion:
a) Tree-level ${\cal O}(p^2)$ diagrams;
b) Tree-level ${\cal O}(p^4)$ diagrams.  The
dashed lines represent longitudinally-polarized
gauge bosons; the wavy lines denote their
transversly polarized partners.  The heavy dots
represent vertices from Eq.~\nela, and the scale
$\Lambda \lsim 4 \pi v$.}

\FIG\ftvs{One-loop ${\cal O} (p^4)$ diagrams from Eq.~\lola\
that contribute to $V_L V_L$ pair production via vector
boson fusion.  The terms of enhanced electroweak strength
are those with Goldstone bosons in the loop.}


\chapter{Vector Boson Fusion}

In this section we will consider the production of
longitudinal vector bosons by the process of vector
boson fusion.  Previous studies have examined the
case with two longitudinal vector
bosons in the initial state.  At high energies, this process
dominates the scattering of vector bosons with two transverse
(or one transverse and one longitudinal) polarizations.

\REF\ferus{S.~Dawson and G.~Valencia, {\it Nucl. Phys.}
{\bf B348} (1991) 23.}
This is illustrated in Figs.~\fgbs\ and \ftvs, where we show
the counting rules that isolate the terms of enhanced
electroweak strength.  The counting rules indicate that the
most important diagrams are those with longitudinal
polarizations in the external states.  They also imply
that the most important radiative corrections
come from diagrams with Goldstone bosons in
the loops.\foot{One
might think that contributions from the heavy top quark
would play an important role because they are enhanced
by factors of $M_t^2/M_W^2$.  In Ref.~\ferus\ it was shown
that these contributions are small for the expected values
of the top quark mass. They could be important,
however, for a heavy fourth generation of fermions.}
These are the radiative corrections
of enhanced electroweak strength, as in
standard model calculations
\refmark{\smgbs}.

This power counting is completely correct for final
states, so we will restrict our attention to final
states of purely longitudinal polarization.  For
initial states, however, the power counting is
too naive:  it ignores the fact that the luminosity
of transverse pairs is much larger than that of
longitudinal pairs
\Ref\bjco{S.~Dawson, in R.~Donaldson and J.~Morfin, eds.,
\it Design and Utilization of Superconducting Super
Collider, \rm Fermilab (1985).}.
The large transverse luminosity can
compensate for the relatively smaller subprocess
cross sections.  Therefore we shall study all production
mechanisms, including those with transverse
vector bosons in the initial state.

\section{Tensor structure}

We begin by decomposing the amplitude into five different tensor
forms that contain ``transverse,'' ``longitudinal,'' and
``mixed'' polarization pieces. We ignore the possibility of
epsilon tensors because they do not appear at the order
to which we are working.  Therefore we write:
$$
\eqalign{
M\ =&\ \epsilon_\mu(q_1) \epsilon_\nu(q_2) M_{\mu \nu} \crr
M_{\mu \nu}\ =&\ T_{\mu \nu}\ +\ L_{\mu \nu}\ +\ X_{\mu
\nu}\ ,\cr}
\eqn\tenffd
$$
where we adopt the notation
$$
V_\mu(q_1) V_\nu(q_2)\ \rightarrow\ w (p) w(p^{\prime})\ .
\eqn\notkin
$$
In the final state, we invoke the equivalence theorem
and denote the longitudinal vector particles by their
corresponding would-be Goldstone bosons.\foot{Note
that $M(V_\mu V_\nu \ra ww) = - M(V_\mu V_\nu \ra V_LV_L)$.}
For charged particles, we take $q_1=q^+$ and $p= p^+$.

In Eq.~\tenffd,
$T_{\mu \nu}$ denotes the tensor for
transverse gauge bosons. It satisfies the conditions
$$
T_{\mu \nu}\,q_{1}^{\mu}\ =\ 0 \qquad{\rm and}
\qquad T_{\mu \nu}\,q_{2}^{\nu}\ =\ 0\ ,
\eqn\contr
$$
and contains two form factors
$$
\eqalign{
T_{\mu \nu}\ =&\ T_s(s,t,u)\bigg(-{s\over 2} \gm
+q_{2\mu}q_{1\nu}
\bigg)\crr
+&\ T_d(s,t,u)\bigg({ut\over 2} \gm\ +\ up_\mu q_{1\nu}\ +\
sp_\mu p_\nu\ +\ tq_{2\mu}p_\nu\bigg)\cr}
\eqn\transffac
$$
When the initial or final states contain identical
particles, Bose symmetry requires that
$$
\eqalign{
T_s(s,t,u)\ &=\ T_s(s,u,t) \crr
T_d(s,t,u)\ &=\ T_d(s,u,t)\ .\cr}
\eqn\idetpa
$$
Note that gauge invariance implies that amplitudes
involving two photons must reduce to this form.
Eq. \transffac\ is also, of course, the form that is found
for the $gg \ra V_L V_L$ amplitudes in the gluon
fusion process.

In Eq.~\tenffd, $L_{\mu \nu}$ denotes the part of the
amplitude that describes the purely longitudinal
polarizations.  In particular, this
amplitude must vanish when contracted with a transverse
polarization vector, which implies
$$
L_{\mu \nu}\ =\ T_l(s,t,u)\, q_{1\nu}q_{2\mu}\ .
\eqn\longff
$$
Then the $V_L V_L \ra V_L V_L$ amplitudes are given
by
$$
-\ {s^2 \over 4 M_{V_1} M_{V_2}} T_l (s,t,u)\ .
\eqn\etch
$$
We have checked our results by comparing with the amplitudes
computed directly from the equivalence theorem.

The remaining tensor structure corresponds to states of
mixed polarization, with one transverse and one longitudinal
vector boson.  There are two form factors for this part
of the amplitude:
$$
\eqalign{
X_{\mu \nu}\ &=\  T_{m1}(s,t,u)\biggl(\tone\biggr) \crr
&+\ T_{m2}(s,t,u)\biggl(\ttwo\biggr)\ .\cr}
\eqn\mixff
$$
As above, we have checked our results by contracting with
one longitudinal polarization vector and comparing with the
results obtained from the equivalence theorem.

\section{Independent Amplitudes}

Since our chiral Lagrangian preserves an approximate $SU(2)$
symmetry, we can use isospin arguments to reduce the number
of independent form factors.  In this section we work in
the $(W_3, B)$ basis because it simplifies the isospin
properties of the vector bosons.

For amplitudes involving two $W$'s in the initial state,
all the particles are isospin triplets.  The reaction is
then characterized by four isovector indices $i\; j
\rightarrow
k\;l$.  Ignoring the tensor structure (but remembering
that interchanges of $s,t,u$ also imply interchanges
of momenta and Lorentz indices), we write:
$$
M^{ijkl}(s,t,u)\ =\ A(s,t,u)\,\delta^{ij}\delta^{kl}
\ +\ B(s,t,u)\,\delta^{ik}\delta^{jl}
\ +\ C(s,t,u)\,\delta^{il}\delta^{jk} \ .
\eqn\isott
$$

Since the particles in the initial and final states are
typically not the same, we can only use $t
\leftrightarrow u$ crossing to simplify this
expression.  We find
$$
\eqalign{
A(s,t,u)\ =&\ A(s,u,t)\cr
C(s,t,u)\ =&\ B(s,u,t)\ ,\cr}
\eqn\bosett
$$
which implies:
$$
M^{ijkl}(s,t,u)\ = \ A(s,t,u)\delta^{ij}\delta^{kl}
\ +\ B(s,t,u)\delta^{ik}\delta^{jl}
\ +\ B(s,u,t)\delta^{il}\delta^{jk} \ .
\eqn\isorestt
$$

\REF\cgg{M.~Chanowitz,
H.~Georgi and M.~Golden, {\it Phys. Rev.} {\bf D36} (1987)
1490; {\it Phys. Rev. Lett.} {\bf 57} (1966) 616.}

We see that there are two independent functions for
amplitudes involving only the $W$ bosons.  They are:
$$
\eqalign{
M(W^+ W^- \ra z z)\ =&\ A(s,t,u)\cr
M(W^+ W^3 \ra w^+ z)\ =&\ B(s,t,u)\ .\cr}
\eqn\expfftt
$$
The other $W$ amplitudes can then be reconstructed from the
relations,
$$
\eqalign{
M(W^3 W^3 \ra w^+ w^-)\ =&\ A(s,t,u)\cr
M(W^- W^3 \ra w^- z)\ =&\ B(s,t,u)\cr
M(W^+ W^- \ra w^+ w^-)\ =&\ A(s,t,u)\ +\ B(s,t,u)\cr
M(W^\pm W^\pm \ra w^\pm w^\pm)\ =&\ B(s,t,u)\ +\ B(s,u,t)\cr
M(W^3 W^3 \ra z z)\ =&\ A(s,t,u)\ +\ B(s,t,u)\ +\ B(s,u,t)\
.\cr}
\eqn\recampff
$$
Note that for purely longitudinal scattering, $s
\leftrightarrow t$ crossing implies $B(s,t,u) = A(t,s,u)$.

The amplitudes involving two hypercharge bosons are a-priori
independent.  From our Lagrangian, it is not hard to see
that
$$
\eqalign{
M(B B \ra z z)\ =&\ \tq^2 M(W^3 W^3 \ra z z) \cr
M(B B \ra w^+ w^-)\ =&\ \tq^2 M(W^3 W^3 \ra w^+ w^-)
\ . \cr}
\eqn\relforbb
$$
The amplitudes involving one $W$ and one $B$ cannot be simplified
with isospin arguments and we must compute them explicitly.

In the appendix, we present explicit results for the
independent scattering amplitudes.  The physical
amplitudes can be reconstructed using the relationships
of this and the previous section.

\chapter{Quark Anti-quark Annihilation}

\REF\ehlq{E.~Eichten,
I.~Hinchliffe, K.~Lane and C.~Quigg, {\it Phys. Rev.} {\bf
D34} (1986) 1547.}

Light $q \overline{q}$ annihilation is the most important
mechanism for vector boson pair production
in hadronic colliders
\REF\dkr{M.~Duncan, G.~Kane and W.~Repko, {\it Nucl. Phys.}
{\bf B272} (1986) 833.}
\refmark{\ehlq,\dkr}.  This process tends to produce
transversely-polarized vector bosons and has been typically
considered as a
background to new physics.  In traditional studies of a
heavy standard-model Higgs boson, one tries to suppress
this mechanism with appropriate cuts
\Ref\han{V.~Barger, K.~Cheung, T.~Han,
J.~Ohnemus, D.~Zeppenfeld,
{\it Phys. Rev.} {\bf D44} (1991) 1426.}.

Quark anti-quark annihilation also produces
a smaller number of longitudinal vector boson
pairs in an $I=1$ state (to the extent that
quark masses can be ignored).  This
production mechanism must be considered when
searching for
\REF\ahn{C.~Ahn, M.~Peskin, B.~Lynn and S.~Selipsky, {\it
Nucl. Phys.} {\bf B309} (1988) 221.}
new physics with isotriplet
resonances like the techni-rho \refmark{\hgb,\ahn}.

For our calculation we assume standard couplings of the
gauge bosons to the light fermions.  Nonetheless, the
next-to-leading terms in the chiral Lagrangian \nela\ affect
the production of $V_LV_L$ pairs through $q \overline{q}$
annihilation.
This has been discussed in Refs.~\wwcpt\ and \simm.
In Ref.~\wwcpt\ an estimate of the rescattering
of the $V_LV_L$ pair was made by considering its absorptive
part.  Since the $V_LV_L$ pair is produced in an $I=1$
state, the
rescattering is sensitive to $L_{1,2}^r(\mu)$ through the
$I=1, J=1$ partial wave.  This effect is ${\cal O}(p^6)$
in the energy expansion.

At order $p^4$, $q \overline{q}$ annihilation
is sensitive to the parameters $L_{9L,R}^r(\mu)$, as
discussed\foot{It is also sensitive to $L^r_{10}(\mu)$,
but not through terms of enhanced electroweak strength.}
in  Ref. \simm.
Our calculations differ from those of Ref. \simm\ in two ways.
First, the authors of Ref. \simm\ did not include loop effects,
which enter the amplitude at the same order as the
$L_{9L,R}^r(\mu)$.  Second, we did not compute any
${\cal O}(g^4)$ contributions, which are suppressed
by $M^2_W/s$ with respect to the leading terms.
In our calculations, we include the full set of
{\cal O}($p^4$) terms.  These contributions are of enhanced
electroweak strength; they dominate the scattering amplitudes
at high energies.

\FIG\fqqan{Counting rules for the production of $V_LV_L$
pairs through $q\overline{q}$ annihilation:  a)
Tree and one-loop amplitudes from Eq.~\lola. The diagrams with
Goldstone bosons in the loop are enhanced at high energies;
b) Tree-level terms from Eq.~\nela; c) Possible new physics
contributions that are not included.  The first
are oblique corrections that are not of
enhanced electroweak strength.  The second are
nonstandard fermion couplings that are excluded
by assumption.}

We obtain the terms of enhanced electroweak strength from
the diagrams depicted schematically in Fig.~\fqqan.  If we
define form factors for the $\gamma w^+ w^-$ and $Z w^+ w^-$
vertices by
$$
\eqalign{
A(\gamma\ra w^+ w^-)_{\mu}\ =&\ -e (p^+ -p^-)_\mu f_\gamma(s)
\crr
A(Z\ra w^+ w^-)_{\mu}\ =&\ -{g \over 2 \cos{\theta_W}}
(p^+ -p^-)_\mu \cos(2\theta_W) f_Z(s) \ ,\cr  }
\eqn\defff
$$
we find the following one-loop result in Landau gauge:
$$
\eqalign{
f_\gamma(s)\ =&\ 1 \ -\ {s \over 96 \pi^2 v^2}
{\rm log}\biggl(
-{s \over \mu^2}\biggr)\ +\ {s \over 16 \pi^2 v^2}
\biggl[L^r_{9L}(\mu)+L^r_{9R}(\mu)\biggr] \crr
f_Z(s)\ =&\ f_\gamma(s)\ +\ {1 \over \cos(2\theta_W)} \biggl[
 {s \over 16 \pi^2 v^2}(L^r_{9L}(\mu)-L^r_{9R}(\mu))
\biggr]\ . \cr}
\eqn\pff
$$

In the helicity basis, the amplitude for
$q(k^+,\lambda)\overline{q}(k^-,\lambda^{\prime})
\rightarrow w^+(p) w^-(p^\prime)$
can be written as
$$
A_{\lambda \lambda^{\prime}}\ =\ C_{\lambda
\lambda^{\prime}}\sin{\theta}\ ,
\eqn\qqhamp
$$
where $\theta$ is the angle between $\vec{k}^+$ and
$\vec{p}$ in the
$q\overline{q}$ center of mass frame.  Using Eq.~\pff, and
assuming that
the couplings of the gauge bosons to the light quarks are
the same as
in the standard model, we find
$$
\eqalign{
C_{+-}\ =&\ g^2\biggl[\sin^2(\theta_W)Q_q f_\gamma(s)\ -\
{\sin^2(\theta_W)\cos(2\theta_W) Q_q \over
2 \cos^2(\theta_W)}f_Z(s)\biggr]\crr
C_{-+}\ =&\ -g^2\biggl[\sin^2(\theta_W)Q_q f_\gamma(s)\ -\
{(\sin^2(\theta_W) Q_q- T_3)\cos(2\theta_W) \over
2 \cos^2(\theta_W)}f_Z(s)\biggr]\ ,\cr}
\eqn\qqres
$$
to one-loop order, with $T_3 = \pm 1/2$.
In a similar way, for $u(k^+, \lambda)\overline{d}(k^-,
\lambda^\prime)\rightarrow w^+(p) z(p^\prime)$,
we find
$$
A_{\pm \mp}\ =\ - {g^2 \sin\theta V_{ud} \over 4 \sqrt{2}}
\biggl\{1-{s \over 96 \pi^2 v^2}
\biggl({\rm log}\biggl(
-{s \over \mu^2}\biggr)-12L^r_{9L}(\mu)\biggr)
\biggr\}\ ,
\eqn\forcc
$$
where $V_{ud}$ is the Kobayashi-Maskawa mixing angle.
Note that the amplitude for
$q\overline{q}\rightarrow zz$
vanishes in the limit of massless quarks.

\chapter{Numerical Results}

In this section we present the complete
cross section for the process $pp \ra V_L V_L$,
to order $p^4$ in $SU(2) \times SU(2)$
chiral perturbation theory.  We include the contributions
from all production mechanisms: $q {\overline q}$
annihilation, vector boson fusion and gluon fusion
through a top-quark loop \refmark{\gluus}.  Our results
contain all terms of enhanced electroweak strength, and
are correct to order $p^4$ in chiral perturbation
theory.

For initial states with quarks and antiquarks,
we find the hadronic cross sections in the
usual way, by convoluting the subprocess
cross sections with the EHLQ \refmark{\ehlq}\
structure functions (set 1), evaluated at
$Q^2 = s$, where $s$ is the squared
center of mass energy for the scattering
subprocess.  For initial states with vector
bosons, we use the effective $W$
approximation to find effective luminosities
for the initial particles.  We evaluate the EHLQ
structure functions at $Q^2=M_W^2$,
which has been shown to be a reasonable
approximation for the standard-model Higgs
\Ref\sfuns{M.~Berger and M.~Chanowitz, {\it Phys.
Lett.} {\bf 263B} (1991) 509; {\bf 267B} (1991)
416.}.
We then compute the full $pp$ cross sections by
folding these luminosities with the subprocess cross
sections,
$$
{d \sigma\over d M_{VV}}(pp\rightarrow V_L V_L)={2
M_{VV}\over S}
{d {\cal L} \over d\tau}\bigg|_{pp/VV}
{\hat \sigma} (M_{VV}^2)
\ .
\eqn\sighat
$$

\REF\lontran{G.~Kane and C.~P.~Yuan, {\it Phys.
Rev. } {\bf D40} (1989) 2231.}
\REF\hann{V.~Barger, K.~Cheung, T.~Han and
D.~Zeppenfeld, {\it Phys. Rev.}
{\bf D44} (1991) 2701.}
\REF\wpwp{V.~Barger, K.~Cheung, T.~Han,
R.~Phillips, {\it Phys. Rev.}
{\bf D42} (1990) 3052; D.~Dicus, J.~Gunion, and
R.~Vega, {\it Phys. Lett.} {\bf B258} (1991) 475.}

In this way we find the phenomenological
cross sections for $V_L V_L$ production
at hadronic colliders.  It is important
to note, however, that in
the $W^+W^-$, $ZZ$, and $W^{\pm}Z$
channels, the longitudinal final
states are dominated by configurations
with transverse polarizations.  The
transverse background must be suppressed
if we are to have any hope of observing the
new physics associated with the longitudinal
final states\foot{As
discussed above, new physics also affects
the transverse states.  The
effects are small, however,
because the terms are not
of enhanced electroweak strength.}
\refmark{\han,\sfuns--\wpwp}.

There are several ways this can be done.
One possibility is to separate the
longitudinally-polarized gauge bosons
from transverse background, as has been
studied in Ref.~\refmark{\lontran}.  A second
proposal is to use forward jet tagging to
eliminate the states produced by $q \overline q$
annhilation and gluon fusion
\refmark{\han,\hann,\wpwp}.  A third suggestion
is to study the $W^\pm W^\pm$ channel
because it receives no contribution from
the $q\overline q$ or $gg$ initial states
\refmark{\sfuns,\wpwp}.

\FIG\wpwmloop{$d \sigma/dM_{WW}$ for $p p\rightarrow
W^+_LW^-_L$ at $\sqrt{S}=40$ TeV. The amplitudes include
only the universal contributions at order $p^4$, that
is, we have set $L^r_i(\mu)=0$ at $\mu = 1.5$ TeV.
The long-dashed line is the contribution
from $q {\overline q}$ annihilation.  The solid, dotted,
and dashed lines are the contributions from $V_L V_L$,
$V_L V_T$, and $V_T V_T$ initial states, respectively.
The dot-dashed line is the
total contribution from vector boson scattering.
The vector boson
curves include both the $W^+_LW^-_L$ and $Z_LZ_L$ initial
states.}

\FIG\zzloop{$d \sigma/dM_{ZZ}$ for $p p\rightarrow Z_LZ_L$
at $\sqrt{S}=40$ TeV.  The amplitudes include only the
universal contributions at order $p^4$, that is, we
have set $L^r_i(\mu)=0$ at $\mu = 1.5$ TeV.
The long-dashed line is the contribution
from $gg$ scattering through a top quark loop with $M_{\rm
top}=200$ GeV.  The solid
and dashed lines are the contributions from $V_L V_L$
and $V_T V_T$ initial states, respectively.  The dot-dashed
line is the total contribution from vector boson scattering.
The vector boson curves
include both the $W^+_LW^-_L$ and $Z_LZ_L$ initial states.}

\FIG\wpzloop{$d \sigma/dM_{WZ}$ for $p p\rightarrow
W^+_LZ_L$ at $\sqrt{S}=40$ TeV.  The amplitudes
include only the universal
contributions at order $p^4$, that is, we have set
$L^r_i(\mu)=0$ at $\mu = 1.5$ TeV.
The long-dashed line is the contribution
from $q {\overline q}$ annihilation.  The solid, dotted,
and dashed lines are the contributions from $V_L V_L$,
$V_L V_T$, and $V_T V_T$ initial states, respectively.
The dot-dashed line is the
total contribution from vector boson scattering.}

\FIG\wpwploop{$d \sigma/dM_{WW}$ for $p p\rightarrow
W^+_LW^+_L$ at $\sqrt{S}=40$ TeV.  The amplitudes
include only the universal
contributions at order $p^4$, that is, we have set
$L_i(\mu)=0$ at $\mu = 1.5$ TeV.  The solid
and dashed lines are the contributions from $V_L V_L$
and $V_T V_T$ initial states, respectively.  The
dot-dashed line
is the total contribution from vector boson scattering.}

In Figs.~\wpwmloop\ $-$ \wpwploop, we show the
cross sections for $pp \ra V_L V_L$, using our
${\cal O}(p^4)$ amplitudes with $L^r_i(\mu)=0$,
for $\mu = 1.5$ TeV.  The figures include all
production mechanisms, and describe a universal
background that is always present in theories
with no light resonances.  From the figures
we see that $q {\overline q}$ annihilation provides
the most important contribution to the $W^+_L W^-_L$
and $W_L^\pm Z_L$ final states.\foot{This process
vanishes for the $Z_L Z_L$ and $W^\pm_L W^\pm_L$
channels.}  The $W^+_L W^-_L$ and $Z_L Z_L$ final
states also receive contributions from
gluon fusion through loops of heavy quarks.
The figures indicate that the contribution
to $Z_L Z_L$ is of the same order as that
from vector boson fusion.  We have not
illustrated the $W^+_L W^-_L$ rate here;
a complete calculation for the standard model
shows that it is significantly smaller than
the contribution from $q \overline{q}$
annihilation
\Ref\gvb{D.~Dicus and C.~Kao, {\it Phys. Rev.}
{\bf D43} (1991) 1555.}.

In Figs.~\wpwmloop\ $-$ \wpwploop, we
also show the contributions from all
polarizations of vector boson fusion,
$V_L V_L \rightarrow V_L V_L$, $V_L V_T
\rightarrow V_L V_L$, and $V_T V_T \rightarrow
V_L V_L$ (where the two transverse polarizations
are summed).  At low energy, we have
$$
\sigma (V_T V_T \rightarrow V_L V_L)\ >\
\sigma (V_T V_L \rightarrow V_L V_L)\ >\
\sigma (V_L V_L \rightarrow V_L V_L)\ ,
\eqn\sigrats
$$
which is entirely due to the magnitudes of
the vector boson luminosities.  We see that at
$M_{VV}\simeq  400$ GeV, the
transversely-polarized vector
bosons increase the rate by a
factor of about two; above this energy they
become less important.  At high energy, we have
$$\eqalign{
\sigma (V_L V_L \rightarrow V_L V_L)\ &\simeq\ {s\over
v^4}\crr
\sigma (V_T V_L \rightarrow V_L V_L)\ &\simeq\ {1\over
v^2}\crr
\sigma (V_T V_T \rightarrow V_L V_L)\ &\simeq\ {1\over s}\
,\cr}
\eqn\sigscals
$$
where $s$ is the squared
center of mass energy in the vector boson
scattering sub-system.  In this regime, the
dominant contribution comes from initial states
with longitudinally-polarized particles.
Our results indicate that the cross-over occurs
at approximately 400 GeV.  Note that the
$V_LV_T$ cross sections for $W^+ W^+ $ and
$ZZ$ vanish at ${\cal O}(p^2)$, so they do
not satisfy Eq.~\sigscals.

\FIG\wpwmco{The total vector boson scattering contribution
to $d \sigma/dM_{WW}$ for $p p\rightarrow W^+_LW^-_L$.
The upper curves have $\sqrt{S}=40$ TeV, while the
lower curves have
$\sqrt{S}=17$ TeV. The solid curves have
all $L^r_i(\mu)=0$ and the dotted curves have
$L^r_1(\mu) = -0.6$,
$L^r_2(\mu)= 0.6$, $L^r_{9L}(\mu)=L^r_{9R}(\mu)=2.4$, and
$L^r_{10}(\mu)=-2.5$, at $\mu=1.5$ TeV.}

\FIG\zzco{The total vector boson scattering contribution to
$d \sigma/dM_{ZZ}$ for $p p\rightarrow Z_LZ_L$.
The upper curves have
$\sqrt{S}=40$ TeV and the lower curves have
$\sqrt{S}=17$ TeV. The solid curves have
all $L^r_i(\mu)=0$ and the dotted curves have
$L^r_1(\mu) = -0.6$,
$L^r_2(\mu)= 0.6$, $L^r_{9L}(\mu)=L^r_{9R}(\mu)=2.4$, and
$L^r_{10}(\mu)=-2.5$, at $\mu=1.5$ TeV.}

\FIG\wpzco{The total vector boson scattering contribution to
$d \sigma/dM_{WZ}$ for $p p\rightarrow W^+_LZ_L$.
The upper curves have
$\sqrt{S}=40$ TeV and the lower curves have
$\sqrt{S}=17$ TeV. The solid curves have
all $L^r_i(\mu)=0$ and the dotted curves have
$L^r_1(\mu) = -0.6$,
$L^r_2(\mu)= 0.6$, $L^r_{9L}(\mu)=L^r_{9R}(\mu)=2.4$, and
$L^r_{10}(\mu)=-2.5$, at $\mu=1.5$ TeV.}

\FIG\wpwpco{The total vector boson scattering contribution
to
$d \sigma/dM_{WW}$ for $p p\rightarrow W^+_LW^+_L$.
The upper curves have
$\sqrt{S}=40$ TeV and the lower curves have
$\sqrt{S}=17$ TeV. The solid curves have
all $L^r_i(\mu)=0$ and the dotted curves have
$L^r_1(\mu) = -0.6$,
$L^r_2(\mu)= 0.6$, $L^r_{9L}(\mu)=L^r_{9R}(\mu)=2.4$, and
$L^r_{10}(\mu)=-2.5$, at $\mu=1.5$ TeV.}

As discussed above,
the major contribution to the $W^+_LW^-_L$ and $W^\pm_L Z_L$
final states comes from $q {\overline q}$ annihilation.
In the $W^+_L W^-_L$ channel, this
is sensitive to $L^r_{9L,R}(\mu)$.
In contrast, the $q {\overline q}$ contribution to the $W^\pm_L Z_L$
channel depends on $L^r_{9L}(\mu)$ only \refmark{\simm}.  The
$Z_L Z_L$ and $W^\pm_L W^\pm_L$ final states do not
receive contributions from $q {\overline q}$ annihilation.
They probe new physics through vector boson
fusion, which depends primarily on $L^r_1(\mu)$ and $L^r_2(\mu)$.
In the $Z_LZ_L$ case, however, the cross section must be
disentangled from the background from gluon fusion through
a top-quark loop.

In Figs.~\wpwmco\ $-$ \wpwpco, we show the total contribution
of vector boson scattering to $V_L V_L$ production for the
SSC and the LHC.  The solid curves were computed with all
the $L^r_i(\mu)=0$, for $\mu = 1.5$ TeV.  The dotted curves
correspond to $L^r_1(\mu) = -0.6$, $L^r_2(\mu)= 0.6$,
$L^r_{9L}(\mu)=L^r_{9R}(\mu)=2.4$, and $L^r_{10}(\mu)=-2.5$,
which result from a spin-one, isospin-one resonance of mass
1.5 TeV.  From the figures we see that the differences
are very small.  Whether or not they can be detected is a
question that requires a full phenomenological analysis of
potential signals, backgrounds and cuts, which is
far beyond the scope of this paper.  In what follows,
we will attempt to make an initial rough estimate,
and leave a more detailed analysis to later work.

\FIG\sqqwz{The number of $W^+_L Z_L$ events per year
with $0.5 < M_{WZ} < 1.0$ TeV, as a function of
$L^r_{9L}(\mu)$ with $\mu = 1.5$ TeV, assuming
an integrated luminosity of $10^{40}$ cm$^{-2}$
at the SSC (solid line) and LHC (dotted line).
With no anomalous couplings, the total number of
$W^+ Z$ events per year is expected to be 88,500
at the SSC and 34,400 at the LHC.  Of these, the
number of $W^+_L Z_L$ events is about 10,000 for
the SSC and 4,000 for the LHC.}

In Fig.~\sqqwz \ we plot the total rate of $W^+_L Z_L$
pairs, integrated over the region $0.5 < M_{WZ} <1.0$
TeV, as a function of $L^r_{9L}(\mu)$ with $\mu = 1.5$
TeV.  Since this channel is particularly sensitive to
$L^r_{9L}(\mu)$, we have set all the other coefficients
$L^r_i(\mu)=0$.  Assuming that it will be possible to
measure the polarization of the final state, and defining
$L^r_{9L}(\mu)$ as being observable if it induces a 50\%
change in the integrated cross section, we see that
the SSC and LHC will be sensitive\foot{The number
of events is much larger at the SSC than the LHC,
however, so the statistical significance of the
results will be larger at the SSC.}
to $L^r_{9L}(\mu)\lsim -3.5$ and $L^r_{9L}(\mu)\gsim
2.5$. If we assume that the polarization measurement
is not possible, the change in the rate
is always less than about $5\%$.  Our results are
consistent with those of Ref.~\simm, and
indicate that polarization measurements will
be necessary for the SSC and LHC to place
meaningful constraints on the three-gauge-boson
vertices $L^r_{9L}(\mu)$ and $L^r_{9R}(\mu)$.

\FIG\sqqww{The number of $W^+_L W^-_L$ events per year
with $0.5 < M_{WW} < 1.0$ TeV, as a function of
$L^r_{9L}(\mu)
= L^r_{9R}(\mu)$ with $\mu = 1.5$ TeV, assuming
an integrated luminosity of $10^{40}$ cm$^{-2}$
at the SSC (solid line) and LHC (dotted line).
With no anomalous couplings, the total number of
$W^+ Z$ events per year is expected to be 174,000
at the SSC and 60,000 at the LHC.}

In Fig.~\sqqww \ we plot the total rate of $W^+_L W^-_L$
pairs, integrated over the region $0.5 < M_{WZ} <1.0$
TeV, as a function of $L^r_{9L}(\mu)
= L^r_{9R}(\mu)$ with $\mu = 1.5$
TeV.  All other coefficients have been set to zero.
Using the above assumptions, we estimate the SSC and LHC
will be sensitive to $L^r_{9}(\mu)\lsim -4.0$ and
$L^r_{9}(\mu)\gsim 3.0$, modulo the question of
backgrounds.  The results are similar to those of
Fig.~\sqqwz.

\FIG\swwww{The number of $W^+_L W^+_L$ events per year
with $0.5 < M_{WW} < 1.0$ TeV, as a function of
$L^r_{1}(\mu)$, with $L^r_2(\mu)=0$ and $\mu = 1.5$ TeV,
assuming an integrated luminosity of $10^{40}$ cm$^{-2}$
at the SSC (solid line) and LHC (dotted line).
The values of $L^r_{1}(\mu)$
preserve unitarity up to 1.0 TeV.}

\FIG\swwwww{The number of $W^+_L W^+_L$ events per year
with $0.5 < M_{WW} < 1.0$ TeV, as a function of
$L^r_{1}(\mu)$, with $L^r_{1}(\mu)= -L^r_2(\mu)=0$
and $\mu = 1.5$ TeV,
assuming an integrated luminosity of $10^{40}$ cm$^{-2}$
at the SSC (solid line) and LHC (dotted line).
The values of $L^r_{1}(\mu)$
preserve unitarity up to 1.0 TeV.}

In Fig.~\swwww\ we plot the total rate of
$W^+_L W^+_L$ pairs in the range  $0.5 < M_{WW}
<1.0$ TeV, as a function of $L^r_{1}(\mu)$,
with $L^r_2(\mu)=0$ and $\mu = 1.5$ TeV.
The values of $L^r_1(\mu)$ are those that
preserve unitarity up to
1~TeV (see Fig.~1).  We have set all the
other coefficients $L^r_i(\mu)=0$.
With the previous assumptions, we
estimate that the SSC and LHC will
be sensitive to $L^r_1(\mu) \lsim -0.75$.
A similar figure with $L^r_1(\mu) = -
L^r_2(\mu)$ in shown in Fig.~\swwwww.
The corresponding limits are $-4.0
\lsim L^r_1(\mu) \lsim  -1.0$ and
$L^r_1(\mu)\gsim 0.8$.
The SSC and LHC will be the first
machines to seriously constrain the
four-gauge-boson vertices
$L^r_1(\mu)$ and $L^r_2(\mu)$.

Of course, we are well aware that
searching for new physics by measuring
deviations in absolute rates requires
a firm understanding of the uncertainties
involved.  Our estimates are subject to
substantial corrections because of detection
issues that we have ignored.  In addition,
we have not addressed the dependence of our
results on the choice of structure functions
or on the scale at which the structure functions
are evaluated.  Nevertheless, the calculations
presented here provide a consistent framework
for more detailed phenomenological studies
of electroweak symmetry breaking.  Clearly,
more work is required before definitive answers
can be found.

\chapter{Summary and Conclusions}

In this paper we computed the cross section
for producing longitudinal vector boson pairs at
hadron colliders.  We included all production mechanisms
(except $g g \rightarrow W_L^+ W_L^-$), and
evaluated our amplitudes to next-to-leading
order in $SU(2) \times SU(2)$
chiral perturbation theory.  The
formalism describes electroweak symmetry
breaking in terms of five free parameters,
and is appropriate for theories with no new
resonances below the TeV scale.  Our
results should give a satisfactory representation
of the scattering amplitudes below one TeV.  Since
they were obtained using the equivalence theorem,
they are valid {\it only} to leading order in
the electroweak gauge couplings.  Our amplitudes
contain the leading corrections from terms of
enhanced electroweak strength.

In the $W^+W^-$, $ZZ$ and $W^\pm Z$ channels,
we saw that $q\overline{q}$ annihilation gives the
most important contribution to the hadronic
cross section.  The rate is dominated by
the transverse modes, whose dependence
on new physics is not of enhanced electroweak
strength.  Clearly, if we wish to study
the physics of electroweak symmetry breaking,
we must isolate the longitudinal final states.
Even in this case, there is an important
model-independent background from gluon
fusion through heavy quark loops.  This
provides a major contribution to the $W^+_L
W^-_L$ and $Z_LZ_L$ channels.\foot{To
${\cal O}(p^4)$, the
$gg \ra Z_LZ_L$ and $gg \ra W^+_L W^-_L$
cross sections are sensitive to anomalous
top couplings
\Ref\petop{R.~Peccei and X.~Zhang, {\it Nucl. Phys.} {\bf
B337} (1990)
269; R.~Peccei, S.~Peris and X.~Zhang, {\it Nucl. Phys.}
{\bf B349} (1991) 305.}.
In the $W^+_L W^-_L$ channel,
the signal is masked by $q\overline{q}$
annihilation.}  In contrast, the $W_L^\pm W_L^\pm$
channel is particularly clean because
it receives no contribution from
$q\overline{q}$ annihilation or gluon
fusion.

\topinsert
\centerline{{\bf Table 3}}
\centerline{Sensitivity of $pp$ colliders to
the physics of electroweak symmetry breaking.}
\vskip.3in
\begintable
\hfil \qquad \hfil |
\hfil \qquad $W^+_LW^-_L$ \qquad \hfil |
\hfil \qquad $Z_LZ_L$ \qquad \hfil |
\hfil \qquad $W^\pm_L Z_L$ \qquad \hfil |
\hfil \qquad $W^\pm_L W^\pm_L$ \qquad \hfil \crthick
$q \overline{q}$ | $L^r_{9L,R}(\mu)$ | 0 | $L^r_{9L}(\mu)$
| 0 \crthick
\ \ $V_LV_L$\ \  | $L^r_{1,2}(\mu)$ | $L^r_{1,2}(\mu)$
| $L^r_{1,2}(\mu)$ | $L^r_{1,2}(\mu)$
\endtable
\bigskip
\vfill
\endinsert

The sensitivity to new physics is shown
in Table 3, where the rows are labelled
by production mechanism, and the columns
by the final state.  Since
$q\overline{q}$ annihilation provides a
significant source of longitudinal
pairs in the $W^+_LW^-_L$ and $W^\pm_L
Z_L$ channels, we
see that these contributions
are sensitive to the parameters
$L^r_{9L}(\mu)$ and $L^r_{9R}(\mu)$
in the effective Lagrangian.
These parameters describe
anomalous three-gauge-boson couplings.

{}From Table 3 we see that $V_LV_L$ fusion
contributes to all final states.  This
process is sensitive to $L^r_{1}(\mu)$
and $L^r_{2}(\mu)$ in the effective
Lagrangian, and probes anomalous
four-gauge-boson couplings.  In the
$W^+_L W^-_L$, $Z_LZ_L$ and
$W^\pm_L Z_L$ channels, the $V_LV_L$
rate must be isolated from the other
production mechanisms.  In the $W^\pm_L W^\pm_L$
channel, however, $V_LV_L$ fusion
gives the most important contribution.
This process can be used to
bound $L^r_1(\mu)$ and $L^r_2(\mu)$.

In this paper we have used chiral perturbation
theory to
explore the mechanism of electroweak symmetry
breaking in the absence of light resonances.
Clearly, our calculations should
be extended to include realistic
backgrounds, cuts and detector
simulations.  Only then can one determine
the full capability of the SSC and LHC
for exploring the physics of electroweak
symmetry breaking.

\centerline{\bf Acknowledgements}

We are grateful to R.~Akhoury, W.~Bardeen,
T.~Han, M.~Veltman and S.~Willenbrock for useful
comments.

\vfill
\eject

\Appendix{A}

In this appendix, we present explicit results for vector
boson scattering into longitudinal gauge-boson pairs.
We work at one-loop order in the
high-energy limit, and assume that the
chiral symmetry group is $SU(2) \times SU(2)$,
spontaneously broken to $SU(2)$.  All the relevant
amplitudes can be extracted using the
relationships of sections 3.1 and 3.2.

\noindent
{\bf I.\qquad $W^3_\mu (q_1) W^3_\nu (q_2)\ \ra\ w^+(p) w^-
$}
$$
\eqalign{
T_s(s,t,u)\ =&\ g^2\biggl[{1 \over s}+{1\over 2\pi^2
v^2}\biggl(L^r_1(\mu)+
{tu\over s^2}L^r_2(\mu) +{1 \over 16}\biggr)\biggr]\crr
-&\ {g^2 \over 96 \pi^2 v^2}\biggl[3 \log\biggl(-{s \over
\mu^2}\biggr)
+{u-t\over  s^2}\biggl(t\log\biggl(-{t \over \mu^2}\biggr)
-u\log\biggl(-{u \over \mu^2}\biggr)\biggr)\biggr]\cr}
$$
$$
T_d(s,t,u)\ =\ g^2\biggl[{1 \over tu}+{1 \over 2 \pi^2 v^2
s}L^r_2(\mu)\biggr]
\ -\ {g^2 \over 48 \pi^2 v^2 s}\biggl[\log\biggl(-{t \over
\mu^2}\biggr)
+\log\biggl(-{u \over \mu^2}\biggr)\biggr]
$$
$$
\eqalign{
T_l(s,t,u)\ =&\ -g^2\biggl[{1\over s}+{1 \over 4\pi^2 v^2}
\biggl(2L^r_1(\mu)+L^r_2(\mu)
-{ 2t u \over s^2}L^r_2(\mu)\biggr)\biggr]\crr
+&\ {g^2 \over 96 \pi^2 v^2}\biggl[3\log\biggl(-{s \over
\mu^2}\biggr)
+{u-t\over s^2}\biggl(u\log\biggl(-{u \over \mu^2}\biggr)
-t\log\biggl(-{t \over \mu^2}\biggr) \biggr) \biggr]\cr}
$$
$$
\eqalign{
T_{m1}(s,t,u)\ =&\ {g^2 \over 4 \pi^2v^2}\biggl[4{tu\over
s^2}-
1\biggr]L^r_2(\mu)\crr
+&\ {g^2 \over 48 \pi^2 v^2}\ {u-t\over s^2}\ \biggl[
u\log\biggl(-{u \over \mu^2}\biggr)-t\log\biggl(-{t \over
\mu^2}\biggr)
\biggr] \cr}
$$
$$
\eqalign{
T_{m2}(s,t,u)\ =&\ {g^2 \over 2\pi^2 v^2}\
{t-u\over s^2}\ L^r_2(\mu) \crr
+&\ {g^2 \over 24 \pi^2 v^2 s^2}\biggl[
u\log\biggl(-{u \over \mu^2}\biggr)-t\log\biggl(-{t \over
\mu^2}\biggr)
\biggr]
\cr}
\eqn\proI
$$

\noindent
{\bf II. $W^+_\mu (q_1) W^3_\nu (q_2) \ra w^+(p) z$}
$$
\eqalign{
T_s(s,t,u)\ =&\ -g^2\biggl[{1 \over s}\ +\ {t \over s^2}\ -\
{1\over 4
\pi^2v^2}\biggl({tu\over
s^2}(2L^r_1(\mu)+L^r_2(\mu))+L^r_2(\mu)\crr
+&\ {1 \over 24}
\biggl(1-{u \over s}\biggr)\biggr)\biggr]
\ -\ {g^2\over 96 \pi^2 v^2}\biggl[\biggl(1-{t\over
s}\biggr)\biggl\{
\log\biggl(-{s \over \mu^2}\biggr) \crr
-&\ {u \over s}
\log\biggl(-{u \over \mu^2}\biggr) \biggr\}
\ +\ 3{ut\over s^2}\log\biggl(-
{t \over \mu^2}\biggr) \biggr] \cr}
$$
$$
\eqalign{
T_d(s,t,u)\ =&\ g^2\biggl[{1 \over su}
\ +\ {1 \over 4\pi^2 v^2 s}(2L^r_1(\mu)+L^r_2(\mu))
\biggr]\crr
-&\ {g^2\over 96 \pi^2  v^2 s}\biggl[
\log\biggl(-{u \over \mu^2}\biggr) + 3\log\biggl(-{t \over
\mu^2}\biggr)
\biggr] \cr}
$$
$$
\eqalign{
T_l(s,t,u)\ =&\ -g^2\biggl[{t\over s^2}\ +\ {1 \over 4 \pi^2
v^2
s^2}\biggl(
2t^2 L^r_1(\mu)+(u^2+s^2) L^r_2(\mu)
\biggl)\biggr]\crr
+&\ {g^2 \over 96 \pi^2 v^2}\biggl[ \biggl(1-{u\over
s}\biggr)
\log\biggl(-{s \over \mu^2}\biggr) +3{t^2 \over s^2}
\log\biggl(-{t \over \mu^2}\biggr)\crr
+&\ {u(u-s)\over s^2}
\log\biggl(-{u \over \mu^2}\biggr) \biggr] \cr}
$$
$$
\eqalign{
T_{m1}(s,t,u)\ =&\ g^2\biggl[ {u-t \over s^2}\ +\ {tu \over
2\pi^2v^2s^2}
(2L^r_1(\mu)+L^r_2(\mu)) \crr
+&\ {1\over 4\pi^2v^2s}\biggl(2tL^r_1(\mu)+
uL^r_2(\mu)\biggr)\biggr]
\ +\ {g^2 (t-u)\over 96 \pi^2v^2 s}\biggl[\log\biggl(-{s
\over
\mu^2}\biggr)
\crr
+&\ 3{t\over s}\log\biggl(-{t \over \mu^2}\biggr) -{u \over
s}
\log\biggl(-{u \over \mu^2}\biggr) \biggr]\cr}
$$
$$
\eqalign{
T_{m2}(s,t,u)\ =&\ g^2\biggl[{2 \over s^2}\ +\ {1\over
4\pi^2v^2
s}(-
2L^r_1(\mu)
+L^r_2(\mu))\crr
+&\ {t-u\over
4\pi^2v^2s^2}(2L^r_1(\mu)+L^r_2(\mu))\biggr]\crr
+&\ {g^2\over 96 \pi^2 v^2 s}\biggl[-2\log\biggl(-{s \over
\mu^2}\biggr)
-6{t\over s}\log\biggl(-{t \over \mu^2}\biggr)
+2{u \over s}\log\biggl(-{u \over \mu^2}\biggr) \biggr]
\cr}
\eqn\proIII
$$

\noindent
{\bf III.\qquad $W^3_\mu (q_1) B_\nu (q_2)\ \ra\ w^+(p) w^-
$}
$$
\eqalign{
{\cal M}\ =&\ -{gg^{\prime} \over 8 \pi^2
v^2}\biggl(L^r_{9L}(\mu)+L^r_{9R}(\mu)
        +2L^r_{10}(\mu)
        -{1 \over 2}\biggr)\ts \crr
        +&\ 2{gg^{\prime}\over ut}\tut \crr
        -&\ \tq{\cal M}(W^3_\mu (q_1) W^3_\nu (q_2) \ra
w^+(p) w^-
)\cr}
\eqn\proII
$$
{\bf IV. $W^+_\mu (q_1) B_\nu (q_2) \ra w^+(p) z$}
$$
\eqalign{
{\cal M}\ =&\ {gg^{\prime} \over 16 \pi^2
v^2}\biggl(L^r_{9L}(\mu)+L^r_{9R}(\mu)
        +2L^r_{10}(\mu)
        +{s-u \over 3 s}\biggr)\ts \crr
        +&2\ {gg^{\prime}\over us}\tut \crr
        -&\ \tq{\cal M}(W^+_\mu (q_1) W^3_\nu (q_2) \ra
w^+(p)
z)\cr}
\eqn\proIV
$$
{\bf V. $W^3_\mu (q_1) B_\nu (q_2) \ra z(p) z$}
$$
\eqalign{
{\cal M}\ =& 2{gg^{\prime} \over 16 \pi^2 v^2} \ts \crr
       &  -\ \tq{\cal M}(W^3_\mu (q_1) W^3_\nu (q_2) \ra
z(p) z)\
.\cr}
\eqn\proV
$$

We have checked that the photon amplitudes, $\gamma \gamma
\ra w^+ w^-$, $\gamma \gamma \ra z z$ and $W^+ \gamma \ra
w^+ z$ are all gauge invariant.\foot{We have also extracted
the $\gamma \gamma \ra \pi \pi$ amplitudes from our results
and checked that they reduce to the amplitudes of Bijnens
and Cornet
\Ref\bjco{J.~Bijnens and
F.~Cornet, {\it Nucl. Phys.} {\bf B296} (1988) 557.}
in their $SU(2)$ and chiral limits, with $v \ra f_\pi
\simeq 93$ MeV. We are grateful to J.~Bijnens for
providing us with
the $SU(2)$ limit of his result to check our amplitudes.}

\refout

\figout
\bye